\documentclass{revtex4}
\usepackage{graphicx}
\usepackage{subfigure,psfrag}
\newcommand{\bs}{\mathbf {s}}
\newcommand{\br}{\mathbf {r}}
\newcommand{\bv}{\mathbf {v}}

\begin{document}

\title{The sensitivity of the vortex filament method to different reconnection models}

\author{A.~W.~Baggaley}
\email{a.w.baggaley@ncl.ac.uk}
\affiliation{School of Mathematics and Statistics, University of
Newcastle, Newcastle upon Tyne, NE1 7RU, UK}

\begin{abstract}
We present a detailed analysis on the effect of using different algorithms to model the reconnection of vortices in quantum turbulence, using the thin-filament approach.
We examine differences between four main algorithms for the case of turbulence driven by a counterflow.
In calculating the velocity field we use both the local induction approximation (LIA) and the full Biot-Savart integral.
We show that results of Biot-Savart simulations are not sensitive to the particular reconnection method used, but LIA results are.
\end{abstract}


\maketitle

\section{Introduction}

Turbulence in the quantum, low temperature phase of liquid helium ($^4$He), also known as quantum turbulence \cite{Donnelly,Barenghi-Sergeev,Halperin}, consists of reconnecting quantized vortex filaments, arranged in  a random, disordered tangle.
Due to quantum mechanical constraints, each vortex carries the same fixed circulation  $h/m$, where $h$ is Plack's constant and $m$ is the mass of one atom. 
This quantity is called the quantum of circulation $\kappa$.  
A number of experimental methods have been used to create this form of turbulence, for example agitating superfluid liquid helium with propellers \cite{Tabeling,Roche2007}, forks\cite{Skrbek}, or
grids \cite{Smith1993}; these techniques are also used to create
turbulence in ordinary fluids. 

At finite temperatures superfluid helium is a two fluid system: a viscous normal fluid component coexisting with an inviscid superfluid component.
The superfluid vortices interact with the thermal excitations which make up the normal fluid, thus introducing a mutual friction force between the two fluid components.
This means that turbulence in the quantum fluid can be driven by the flow of the normal fluid or vice-versa.
One particular example is superfluid turbulence driven by a heat flow \cite{Vinen1957,Paoletti2008,Tough}.
Note that this form of turbulence has no classical analogy.

Recently new flow visualization techniques, such  as
tracer particles\cite{Maryland-tracers,VanSciver}, Andreev scattering
\cite{Lancaster-Andreev} and laser-induced fluorescence\cite{Yale}, have added substantially to our knowledge of the nature of quantum turbulence.
Even with these advances in flow visualization numerical simulations will continue to play a crucial role in furthering our knowledge.
Indeed, building on the pioneering work of Schwarz \cite{Schwarz},
numerical simulations have always been important in the field 
\cite{Samuels,Aarts,Bauer,Tsubota2003,Risto,Konda,Kivo,Morris,Kivo2,cascade,tree}, and the recent experimental progress has highlighted their importance 
in interpreting experimental data \cite{Kivotides-PIV,Finne}. 
This article is concerned with verifying that the vortex reconnection procedure used in the popular vortex filament method (VFM) is robust. 
Before we discuss the modelling of reconnections in the VFM, we shall give a brief outline of the VFM method.

In superfluid helium, the vortex core radius 
($a_0 \approx 10^{-8}~\rm cm$) is many orders of magnitude smaller than the
average separation between vortex lines (typically from
$10^{-2}$ to $10^{-4} \rm cm$) or any other relevant length scale in the flow.  Starting from this key observation, Schwarz \cite{Schwarz} modelled vortex lines as spaces curves $\bs=\bs(\xi,t)$ of
infinitesimal thickness, where $t$ is the time and $\xi$ is arc length, using the classical theory of vortex filaments \cite{saffmanbook}.
In the VFM these space curves are numerically discretized by a large, variable number of
points $\bs_i$ ($i=1,\cdots N)$, which hereafter we refer to as vortex points.
In all implementations of the VFM to superfluids the number of vortex points varies in time, as it is desirable to maintain a relatively constant resolution along the filaments by adding or removing points as the length of the filaments changes.
In a recent paper \cite{cascade} we have discussed the various numerical details at length, here it is suffice to say that the resolution along the filaments lies between an upper, $\delta$, and lower, $\delta/2$, bound.

The motion of a vortex filament is determined by the normal fluid, through mutual friction, and by the induced velocity of the other vortices.
The governing equation of motion of the superfluid vortex lines, at point $\bs$ is given by the Schwarz equation \cite{Schwarz}

\begin{equation}
\frac{d\bs}{dt}=\bv_s+\alpha\bs' \times (\bv_n-\bv_s)
-\alpha'\bs' \times \left[ \bs' \times (\bv_n-\bv_s)\right],
\label{eq:Schwarz}
\end{equation}

\noindent
where 
$\alpha$, $\alpha'$ are temperature dependent friction coefficients
\cite{Barenghi1983,Barenghi1998}, 
$\bv_n$ is the normal fluid's velocity, and the velocity $\bv_s$ is 
the velocity field the quantised vortices alone induce.
The prime denotes derivative with respect to arclength, e.g. $\bs'=d\bs/d\xi$. 
Note that, strictly speaking, one should also model the back reaction of the superfluid on the normal component; $\bv_n$ should be the solution of the Navier-Stokes equation modified by mutual friction. 
However, for the sake of simplicity, here we shall simply prescribe the normal flow.

In our system the quantised vortices define the vorticity field; we recover the velocity field $\bv_s$ by numerically solving the Biot-Savart (BS) integral,
\begin{equation}
\bv_s(\bs)=-\frac{\kappa}{4 \pi} \oint_{\cal L} \frac{(\bs-\br) }
{\vert \bs - \br \vert^3}
\times {\bf d}\br,
\label{eq:BS}
\end{equation}
where the line integral extends over the entire vortex configuration
$\cal L$. 
We de-singularize the integral in a standard way \cite{Schwarz}.

Schwarz \cite{Schwarz} proposed the use of the Local Induction Approximation (LIA) \cite{DaRios,Arms-Hama}
as an alternative to the Biot-Savart method, which is computationally very expensive.
The LIA ignores the non-local contribution to the motion of a section of the vortex filament.
Instead a vortex line at the point $\bs_i$ moves along the binormal to $\bs_i$, with the following equation of motion
\begin{equation}
\frac{d\bs_i}{dt}=\frac{\kappa}{4\pi}\ln \biggl( \frac{c R }{a_0} \biggr) \bs_i' \times \bs_i'',
\end{equation}
where $c$ is a constant of order unity and $R$ is the local 
radius of curvature $|\bs_i''|^{-1}$ of the vortex filament. 
We shall test various reconnection algorithms using both the full BS law, and the LIA in a series of numerical experiments.
Again, for further details about spatial derivatives and methods to time-step the vortices, we refer the reader to \cite{cascade}.

We know from experiments \cite{Paoletti2010} and from more microscopic models \cite{Koplik,Tebbs,Kerr,Bajer} that superfluid vortex lines can reconnect with each other when they come sufficiently close, as envisaged by Feynman \cite{Feynman}. 
Superfluid vortex reconnections do not violate Kelvin's theorem as near the axis of the vortex core, where density and pressure vanish and velocity diverges, the governing Gross-Pitaevski equation (GPE) differs from the classical Euler equation.

Whilst vortex reconnections are natural solutions of the GPE,
within the VFM reconnections must be modelled by supplementing Eq.~(\ref{eq:BS})  with an algorithmical reconnection procedure.
This was originally proposed by Schwarz \cite{Schwarz}, and since then a number of alternative algorithms have been proposed.
Until now no detailed test of the effects of varying this procedure has been performed.
Whilst a number of studies have shown good agreement between results using the VFM and experimental results \cite{tree,Adachi2010,Fuji}, a detailed study of the reconnection algorithm is timely; that is the purpose of this study. 

\section{Reconnection algorithms}\label{sec:recon}
\begin{figure*}
  \begin{center}
    \includegraphics[width=0.45\textwidth]{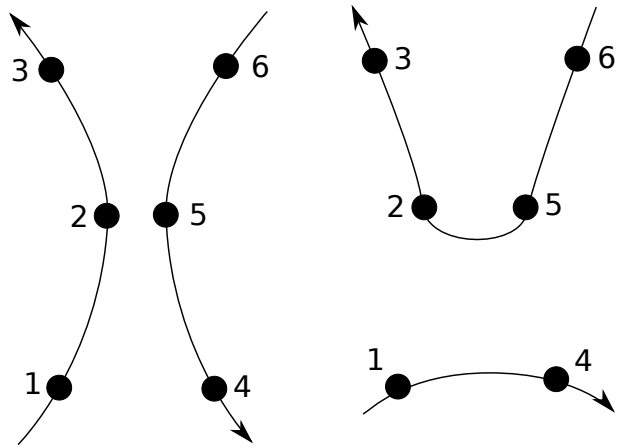}
    \caption{\label{recon1} Schematic reconnection procedure for Type I and II reconnections described in section \ref{sec:recon}. If two vortex points are closer than a critical distance $\Delta=\delta/2$ then a flag swapping operation changes the topology of the filaments and performs the reconnection. From $(1\rightarrow 2\rightarrow 3)$ and $(4\rightarrow 5\rightarrow 6$) to $(6\rightarrow 5\rightarrow 2 \rightarrow 3)$ and $(1\rightarrow 4$).}
  \end{center}
\end{figure*}
In his pioneering paper \cite{Schwarz}, Schwarz's suggested that vortex lines reconnect whenever the distance between a pair of vortices is less than  $\Delta=2R/[c \ln(R/a_0)]$, where $R$ is the radius of curvature at the reconnection point, and $c$ is a constant of order unity.
However, this approach can lead to non-physical reconnections.
Consider, for example, two almost straight vortices; under Schwarz's criterion these vortices must reconnect even if they are very far apart, as a large radius of curvature $R$, results in a large value for $\Delta$.
In this work we therefore avoid the use of this reconnection criterion, instead focusing on methods which have been used in more recent studies.
\begin{figure*}
  \begin{center}
    \includegraphics[width=0.45\textwidth]{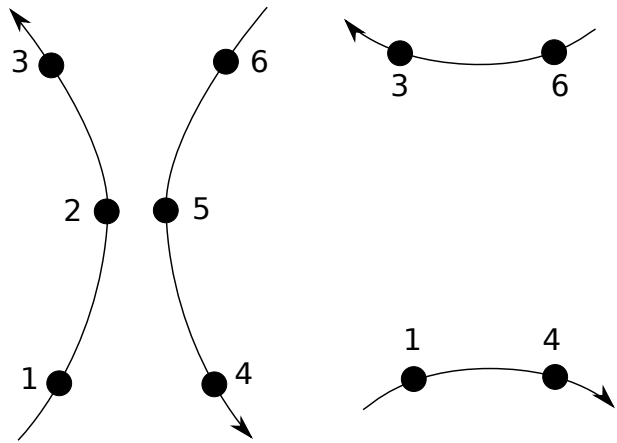}
    \caption{\label{recon2} Schematic reconnection procedure for Type III reconnection algorithm described in section \ref{sec:recon}. The segments $(1\rightarrow 2\rightarrow 3)$ and $(4\rightarrow 5\rightarrow 6$) evolve into $(6 \rightarrow 3)$ and $(1 \rightarrow 4)$. Points 2 and 5 are eliminated.}
  \end{center}
\end{figure*}

Most recent studies have related the reconnection distance to the space resolution along the filaments \cite{Tsubota2000,Tsubota2011}.
We consider three differing reconnection algorithms which take the critical reconnection distance to be related to the maximum spatial resolution, $\Delta=\delta/2$.
We define the first reconnection algorithm as Type I.
In a Type I reconnection vortices are simply reconnected if their separation if less than $\Delta$. 
Motivated by the fact that a reconnection is dissipative event, leading to phonon emission \cite{Leadbeater2001}, we consider an improved algorithm denoted Type II.
As vortex line length is a proxy for the kinetic energy, for a Type II reconnection not only must the distance between the reconnecting filament be less that $\delta/2$, but also the line length must be reduced by the change of topology.
A schematic for the Type I and II algorithms can be seen in Fig.~\ref{recon1}.
The third reconnection model, Type III, can be considered an `ultra' dissipative algorithm.
Type III differs from Type I because the points which triggered the reconnection are eliminated leading to a greater loss of line length, and hence energy.
A schematic of this algorithm is displayed in Fig.~\ref{recon2}.
\begin{figure*}
  \begin{center}
    \includegraphics[width=0.45\textwidth]{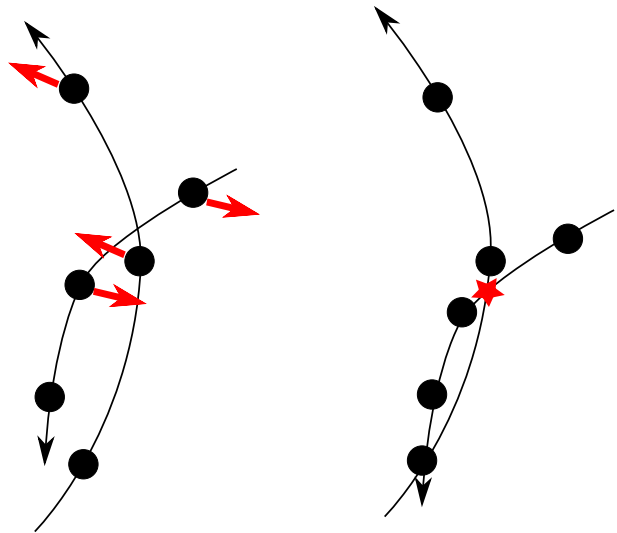}
    \caption{\label{recon3} Schematic representation of the motion of vortex filaments under the Type IV algorithm.
If the segments will collide (as shown) then the reconnection is performed as in Fig.~\ref{recon1}.}
  \end{center}
\end{figure*}

In a recent work Kondaurova and Nemirovskii \cite{Konda} introduced a new model (Type IV) of reconnections within the VFM, which tests whether line segments will meet during the time-step.
Candidate pairs of vortex points ($i$ \& $j$), points which are close enough to reconnect, are identified.
The Type IV algorithm then assumes that the line segments between
each of the pair will move with a constant velocity ($\mathbf{v}(\bs_i)$ \& $\mathbf{v}(\bs_j)$) during
the time-step.
Under this assumption a set of simultaneous equations can be constructed,  
\[
x_i+v_{x}(\bs_i)\Delta t+(x_{i+1}-x_i)\psi=x_j+v_{x}(\bs_j)\Delta t+(x_{j+1}-x_j)\phi
\]
\[
y_i+v_{y}(\bs_i)\Delta t+(y_{i+1}-y_i)\psi=y_j+v_{y}(\bs_j)\Delta t+(y_{j+1}-y_j)\phi
\]
\[
z_i+v_{z}(\bs_i)\Delta t+(z_{i+1}-z_i)\psi=z_j+v_{z}(\bs_j)\Delta t+(z_{j+1}-z_j)\phi
\]
\[
0\leq \phi \leq 1, 0\leq \psi \leq 1
\]
such that if a solution for $\phi$ and $\psi$ can be found, then the filaments will collide during the time-step.
Here $\bs_i=(x_i,y_i,z_i)$, $\bs_{i+1}=(x_{i+1},y_{i+1},z_{i+1})$ $\bs_{j}=(x_j,y_j,z_j)$ and
$\bs_{j+1}=(x_{j+1},y_{j+1},z_{j+1})$ are the coordinates of the pair of points and their neighbours along the filament, and $\Delta t$ is the numerical time-step.
If the line segments will meet, then the filaments are reconnected in the same manner as in Fig.~\ref{recon1}.
We solve for $\psi$ and $\phi$ algebraically using the equations for the  $x$ and $y$ components.
The values of $\psi$ and $\phi$ obtained are then tested in the equation for the $z$ component by verifying that,
\begin{equation}
|z_i+v_{z}(\bs_i)\Delta t+(z_{i+1}-z_i)\psi-z_j+v_{z}(\bs_j)\Delta t+(z_{j+1}-z_j)\phi| < \epsilon,
\end{equation}
where $\epsilon$ is the imposed the tolerance of the method.
We define candidate points ($i$ \& $j$) as vortex points within distance $\Delta=\delta$.
A schematic diagram of the Type IV algorithm is shown in Fig.~\ref{recon3}.

Before performing any reconnection we test the distance from a point $i$ to all other points, which are not the nearest neighbours along the filament.
We then begin the test for a reconnection based on the closest reconnection point, assuming the distance is smaller than $\Delta$.
This means self-reconnections (which can arise if a vortex filament has twisted by a large amount) are treated
in the same manner as reconnections between different filaments.
Finally as reconnections must preserve the orientation of the vorticity,
we check that the two reconnecting filaments are not parallel.
To do this we form local (unit) tangent vectors, for example if the reconnecting points are $i$ and $j$ we can readily calculate $\hat{\bs}_i'$ and $\hat{\bs}_j'$.
We then test that $\hat{\bs}_i' \cdot \hat{\bs}_j' < 0.965$,
which ensures that the minimum angle between reconnecting filaments is approximately $15$ degrees.
We have tested that altering this threshold to smaller angles does not lead to any discernible difference in the results, hence for brevity all results here are subject to this criterion.

\section{Numerical simulations}\label{sec:counterflow}

We now the numerical simulations used to test 
the various reconnection algorithms described in the previous section.
As a benchmark we choose counterflow turbulence, the relative
motion of the normal fluid and superfluid components sustained by an applied heat flow in the direction of $\bv_n$.
The superfluid flows in the opposite direction so that $\rho_n\bv_n+\rho_s\bv_s=\mathbf{0}$, where $\rho_n$ and $\rho_s$ are respectively the normal fluid and superfluid densities.
We choose this form of turbulence as there has been a wealth of 
experimental studies \cite{Vinen1957,Tough,Childers}, as well as some recent detailed numerical simulations \cite{Adachi2010,Adachi2010a}.
Counterflow turbulence was also used in a recent numerical study by Kondaurova and Nemirovskii \cite{Konda2008}, where they reported that the use of a reconnection method, similar to the Type IV algorithm, gave a steady state solution for LIA simulations.
We shall also test these claims in this study.

\begin{figure*}
  \begin{center}
    \includegraphics[width=0.4\textwidth]{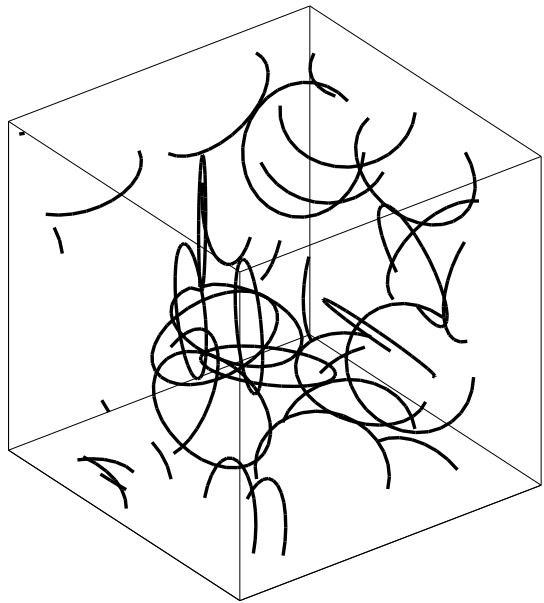}
    \caption{\label{initial_cond} The initial conditions used in every simualtion; a set of random loops with radius $0.0095\,$cm.}
  \end{center}
\end{figure*}

Our calculations are performed in a periodic cube with sides of length 
$D=0.1~\rm cm$. 
Superfluid and normal fluid velocities $\bv_n$ and
$\bv_s$ are imposed in the positive and negative $x$ directions
respectively, where $v_{ns}=\vert \bv_n -\bv_s \vert$ is proportional
to the applied heat flux. 
Simulations are performed with the same numerical resolution
$\delta=1.6 \times 10^{-3}~\rm cm$ and time-step 
$\Delta t=10^{-4}~\rm s$.
The initial condition consists of a set of vortex loops
set at random locations, which are displayed in Fig.~\ref{initial_cond}.

During the evolution we monitor the vortex line density calculated as,
\begin{equation}
\label{eq:Ldensity}
L=\frac{\Lambda}{V},
\end{equation}

\noindent
where $V=D^3$ is the volume of the computational domain, $\mathcal{L}$ 
is the entire vortex configuration and $\Lambda=\int_\mathcal{L} d\xi$ is the total vortex line length.
We also monitor the number of reconnections per unit time (the reconnection rate), $\zeta$.

We run simulations, with the different reconnection algorithms, using both the LIA and the full BS law with $v_{ns}=0.55\,$cm/s and temperature $T=1.6\,$K ($\alpha=0.098$, $\alpha'=0.016$).
We then run further simulations, using only the full BS law, for three further values of $v_{ns}$.
This allows us to test the theoretically predicted \cite{Vinen1957} and experimentally and numerically \cite{Adachi2010} verified law for the steady state line length,
\begin{equation}
L=\gamma^2 v_{ns}^2,
\end{equation}
where $\gamma$ is a temperature-dependent parameter.
Our aim is to determine whether the different reconnection algorithms yield different values of $\gamma$. We stress that $\gamma$ is a macroscopic quantity which is known from experiments.

\section{Results}\label{sec:results}
\begin{figure*}
  \begin{center}
    \psfrag{L}{$L$}
    \psfrag{t}{$t$}
    \includegraphics[width=0.47\textwidth]{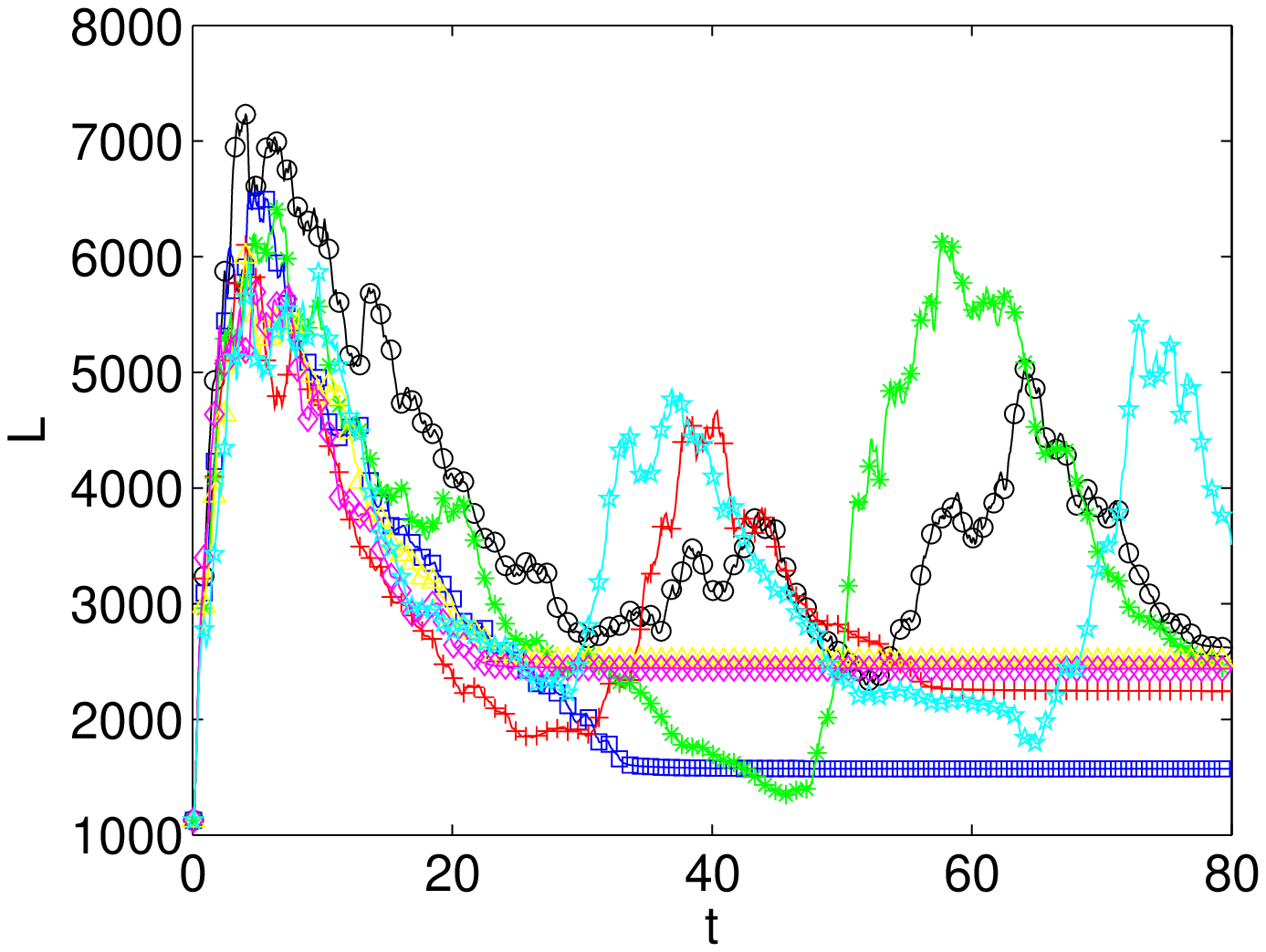}
    \psfrag{L}{$L$}
    \psfrag{t}{$t$}
    \includegraphics[width=0.47\textwidth]{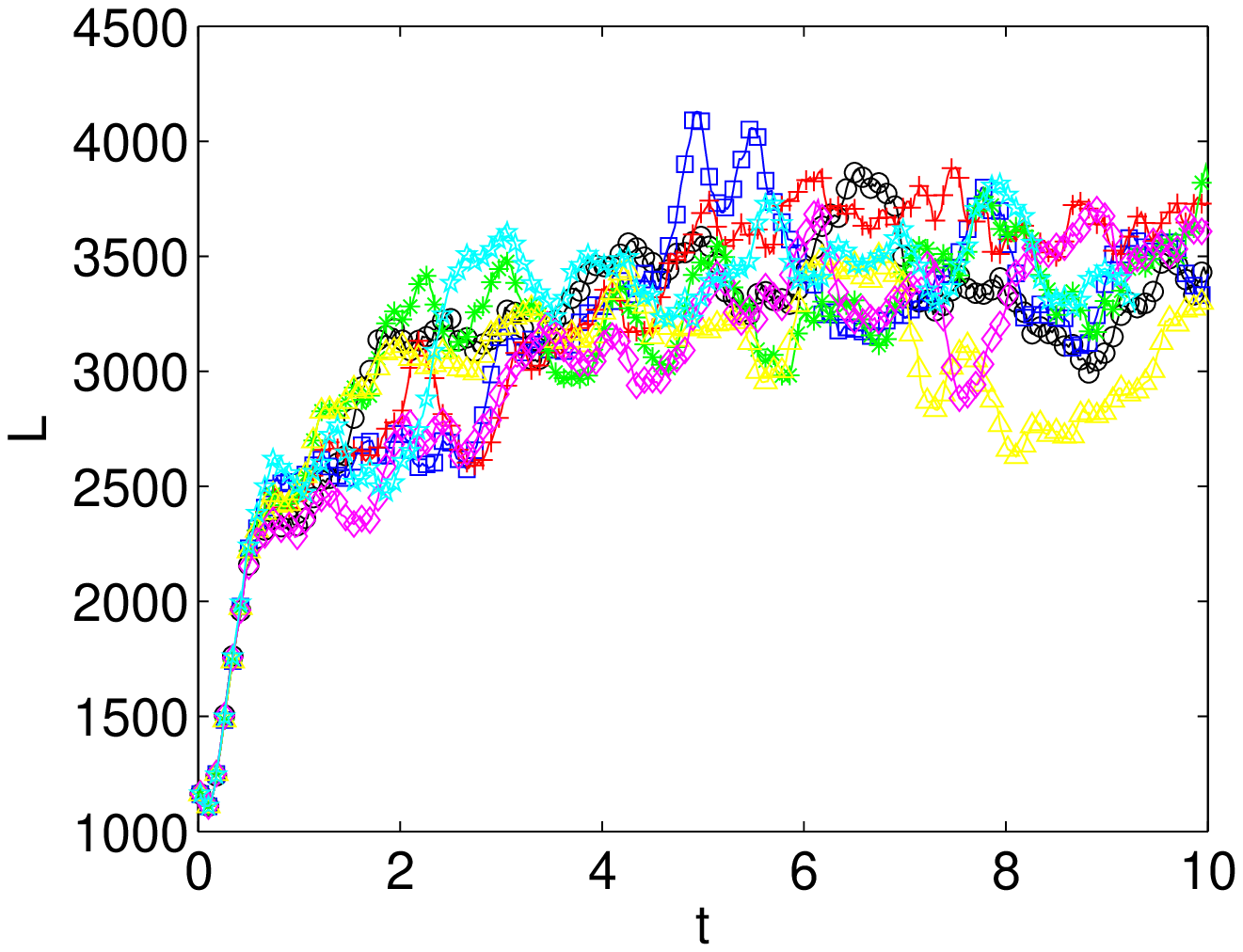}
    \caption{\label{result1} The vortex line density $L$ (cm) plotted as a function of time $t$ (cm) for simulations with the LIA (left) and the full BS law (right) with different reconnection algorithms, $v_{ns}=0.55$cm/s and $T=1.6$K. The symbols are as follows: (magenta) diamonds: Type I reconnection; (yellow) triangles: Type II; (cyan) pentagons: Type III; (black) circles: Type IV ($\epsilon=10^{-4}$); (blue) squares: Type IV ($\epsilon=10^{-3}$); (red) crosses: Type IV ($\epsilon=10^{-2}$) and (green) asterisks: Type IV ($\epsilon=10^{-1}$).}
  \end{center}
\end{figure*}
\begin{figure*}
  \begin{center}
    \psfrag{R}{$\zeta/\Lambda$}
    \psfrag{t}{$t$}
    \includegraphics[width=0.47\textwidth]{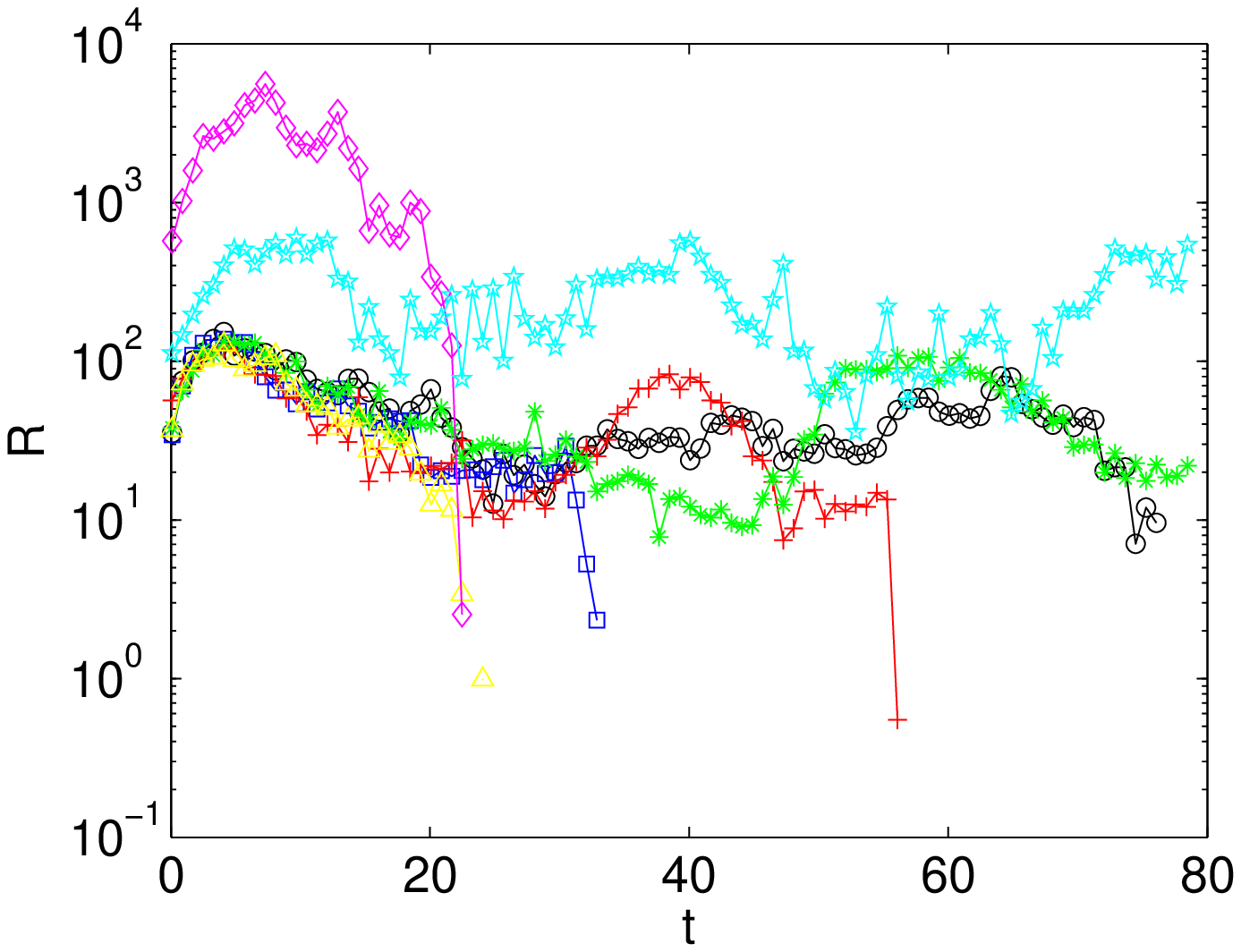}
    \psfrag{R}{$\zeta/\Lambda$}
    \psfrag{t}{$t$}
    \includegraphics[width=0.47\textwidth]{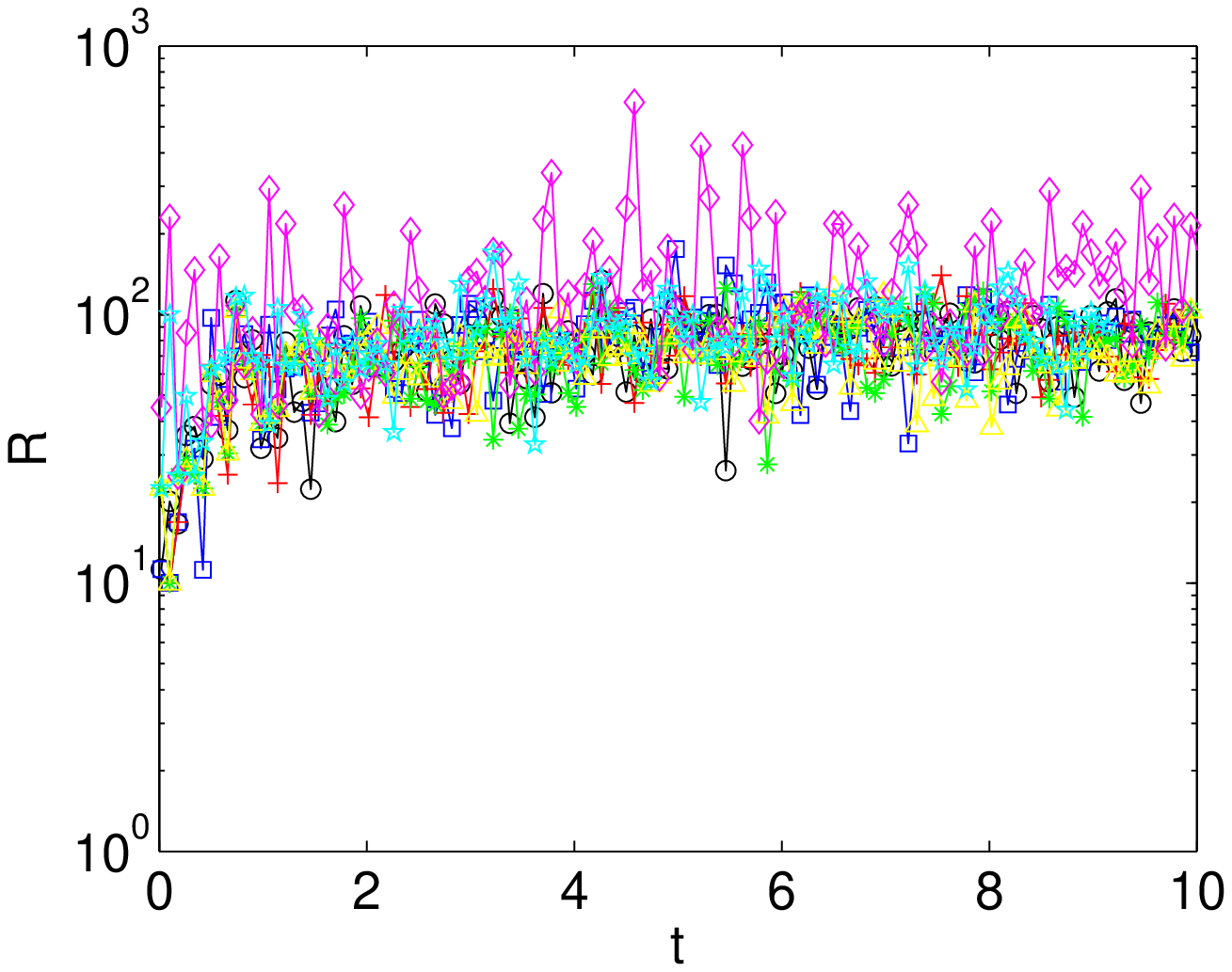}
    \caption{\label{result2} The reconnection rate, $\zeta$ (s$^{-1}$), scaled by the vortex line length, $\Lambda$ (cm), plotted as a function of time, $t$ (s), for the LIA (left) and BS (right) simulations; plotting symbols and colors is as in Fig.~\ref{result1}.
    Note that in the case of the LIA simulations a number of the simulations reach a degenerate steady state, as is visible in Fig.~\ref{result1} (left). 
In this case the reconnection rate drops to zero and hence the results are not visible on this semi-log scale plot. This does not happen in the full BS simulations.}
  \end{center}
\end{figure*}
\begin{figure*}
  \begin{center}
    \includegraphics[width=0.3\textwidth]{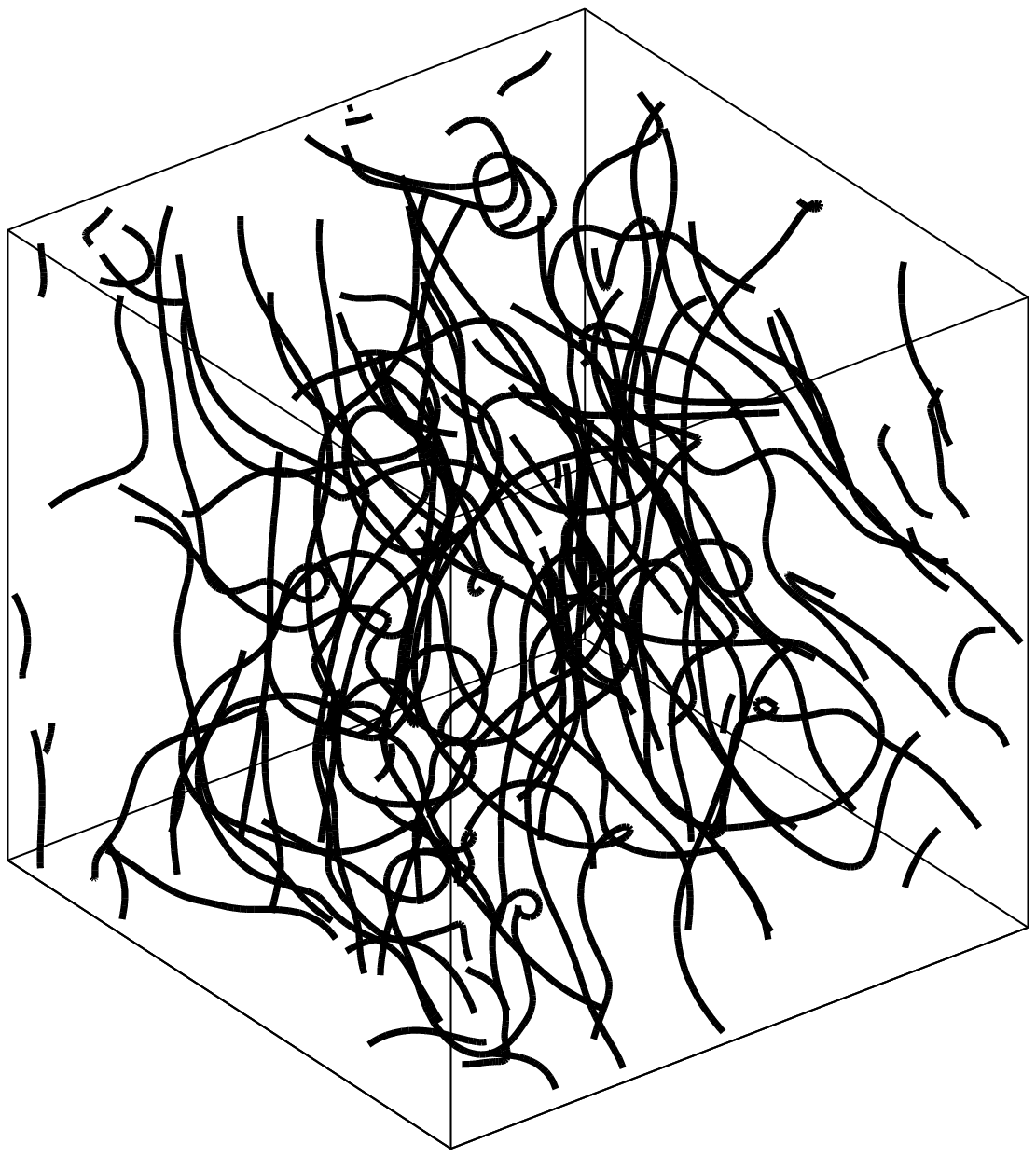}
    \includegraphics[width=0.3\textwidth]{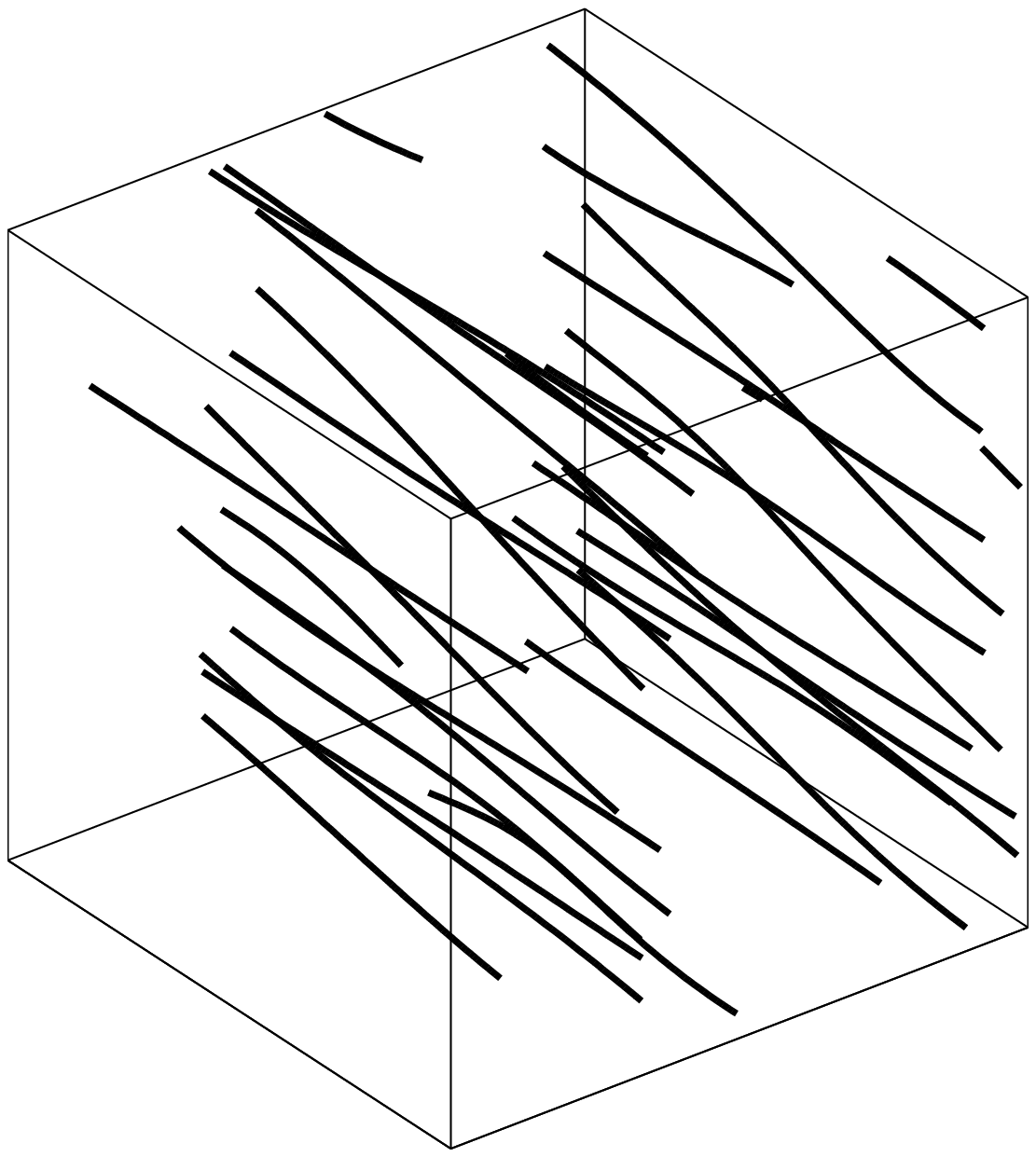}
    \includegraphics[width=0.3\textwidth]{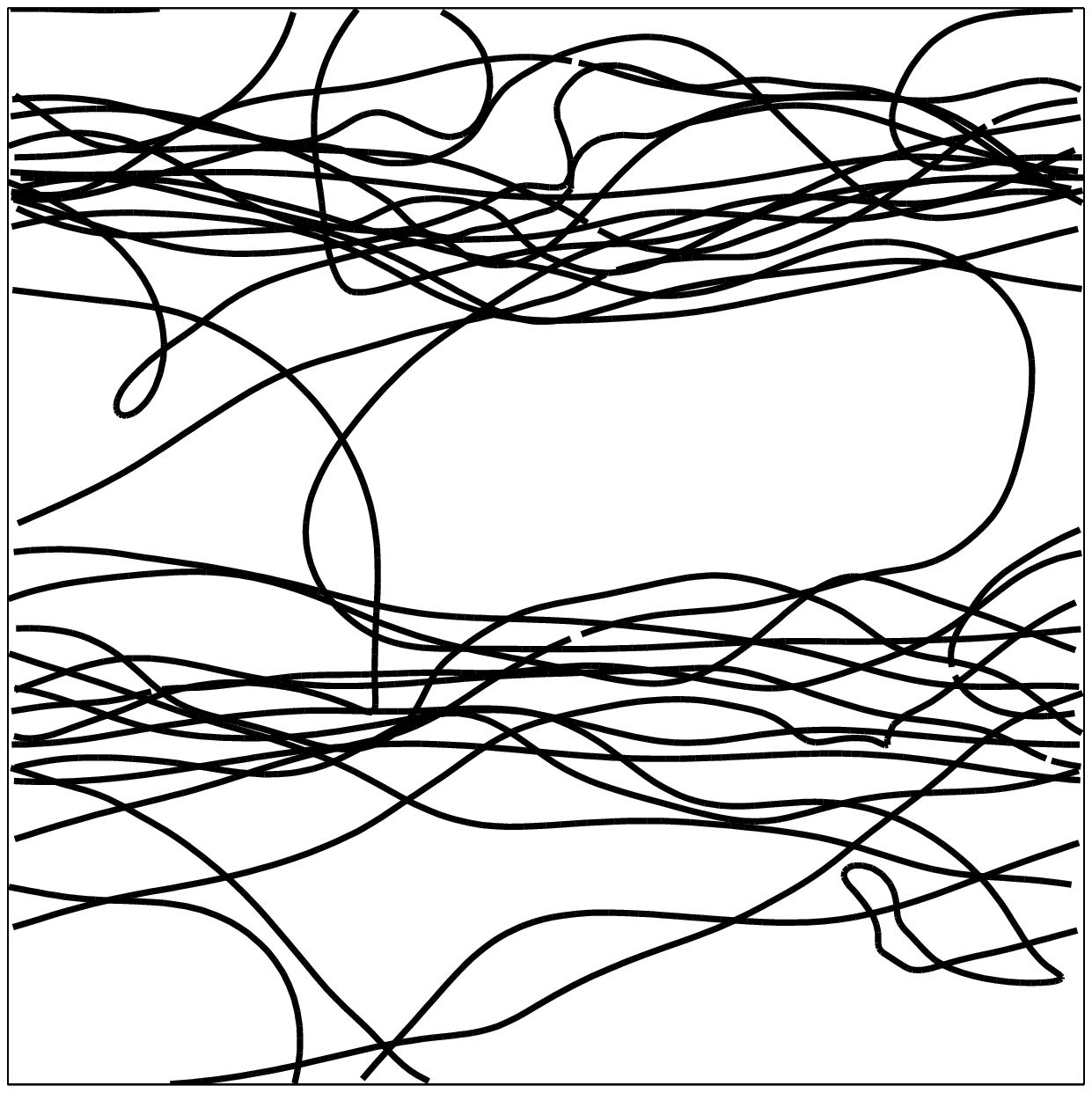}
    \caption{\label{result3} Snapshots of the vortex filaments from the simulations using the LIA, as plotted in Fig.~\ref{result1} (left). 
The left panel shows the typical state of each of the simulations at the peak of vortex line density, at $t \approx 6$(s); this particular plot is taken from the simulation using the Type III reconnection algorithm.
 The middle panel shows the degenerate solution taken at $t=80\,$s for the simulation with Type IV reconnection algorithm ($\epsilon=10^{-3}$). 
The right panel shows a plot in the plane perpendicular to the counterflow for the simulation with the Type IV ($\epsilon=10^{-1}$) algorithm at $t=60\,$s:
this layered vortex structure was recognised by Schwarz \cite{Schwarz} and Adachi et al. \cite{Adachi2010}.}
  \end{center}
\end{figure*}
\begin{figure*}
  \begin{center}
    \includegraphics[width=0.3\textwidth]{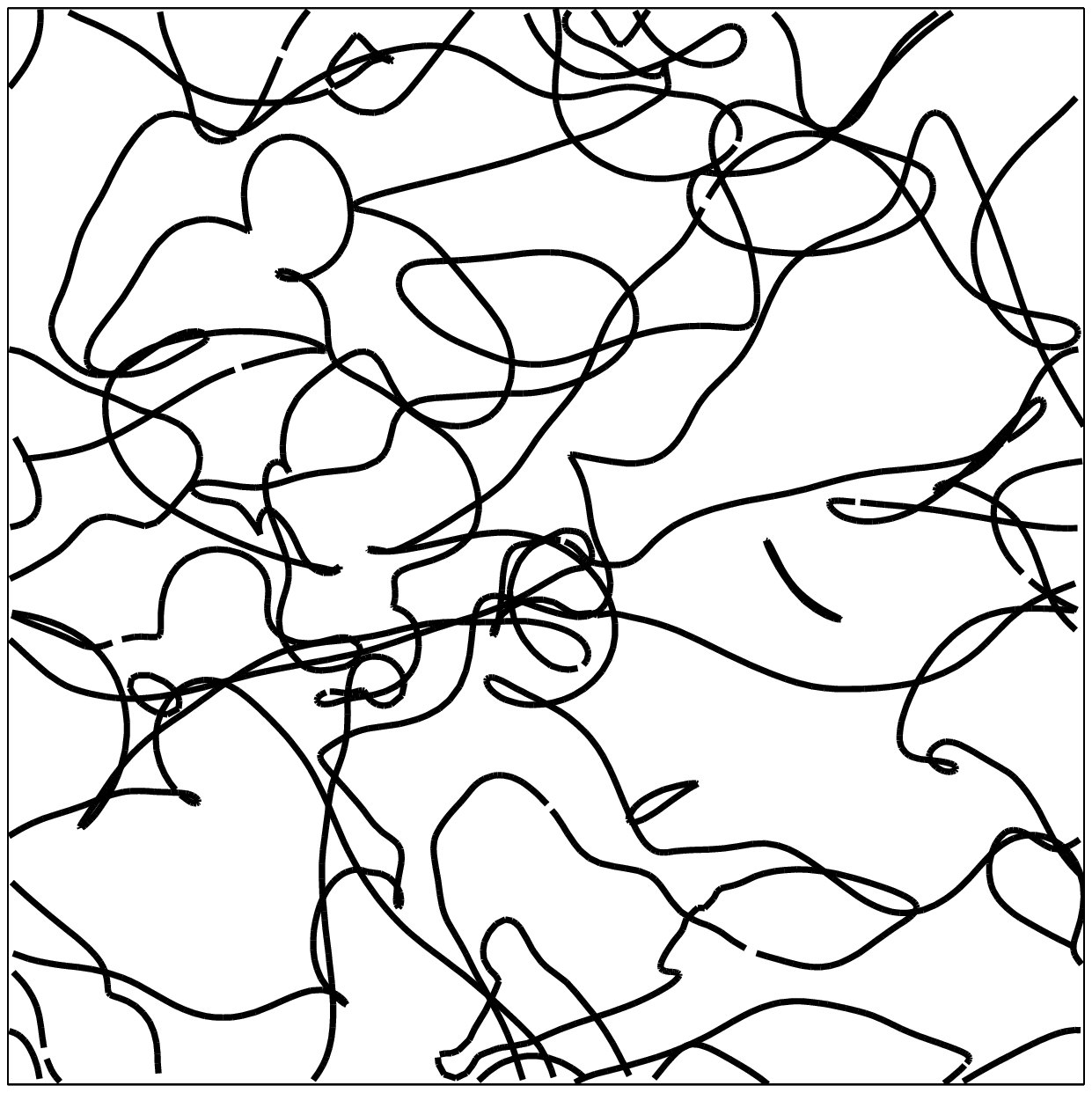}
    \includegraphics[width=0.3\textwidth]{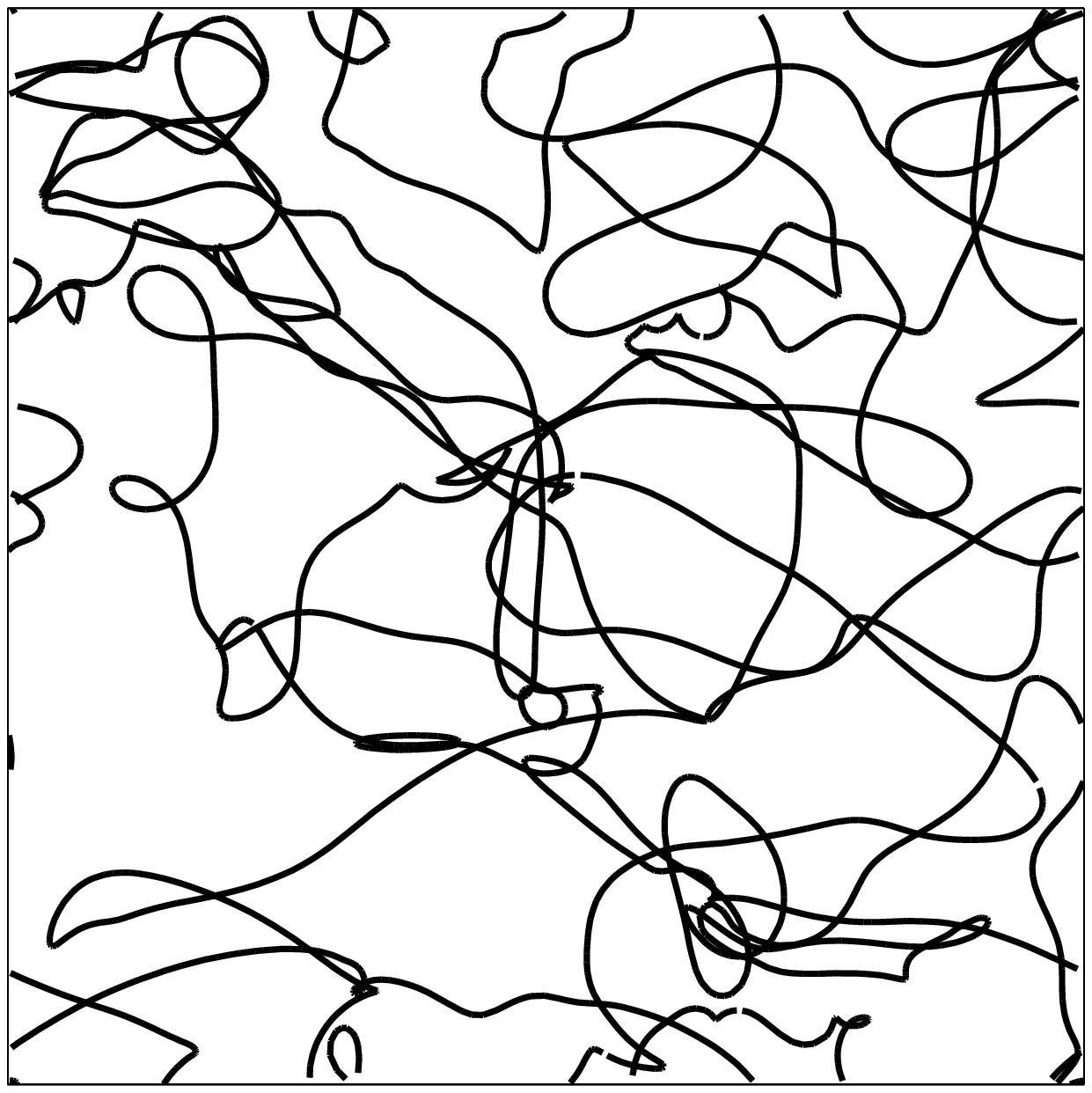}
    \includegraphics[width=0.3\textwidth]{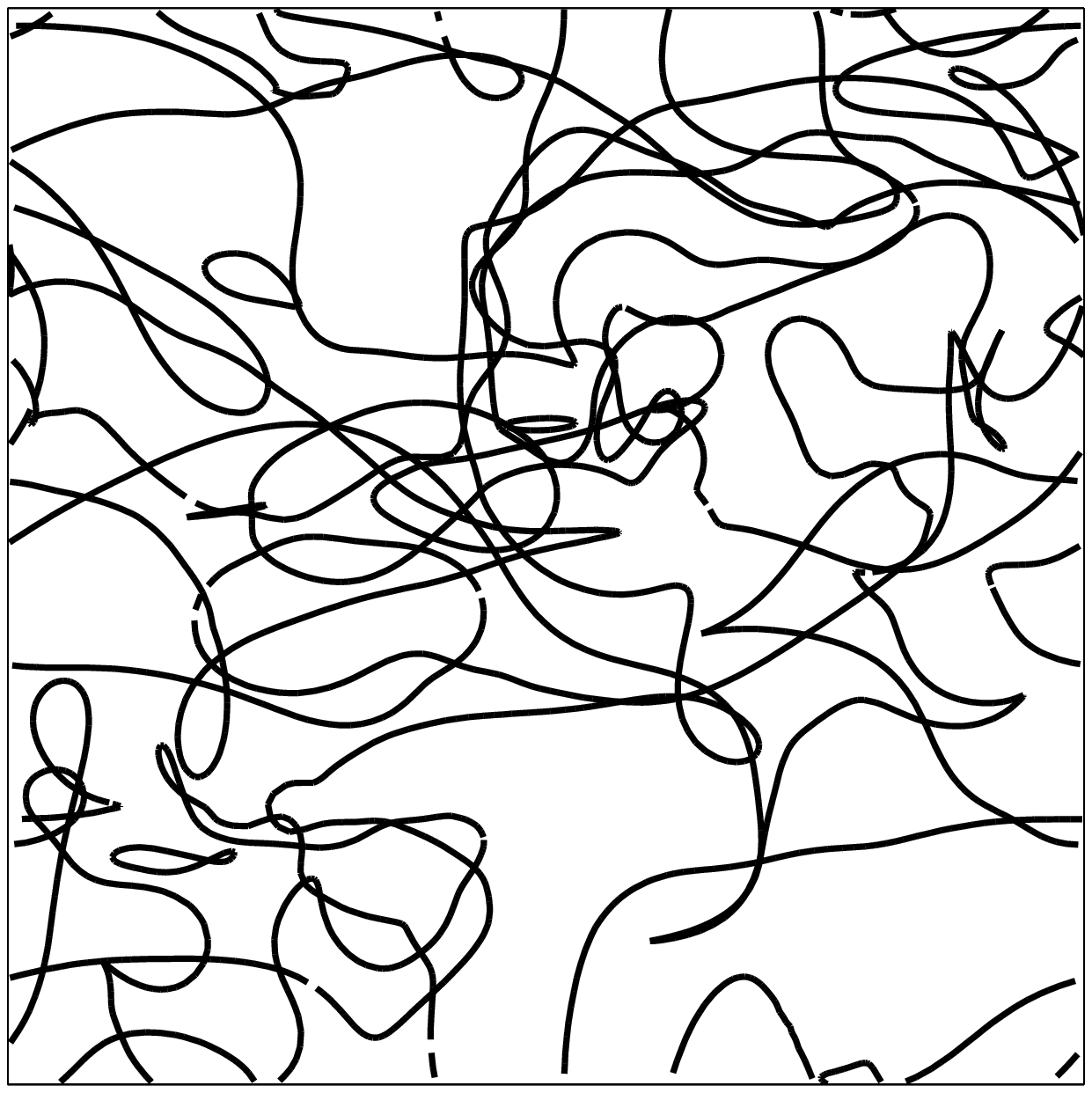}
    \caption{\label{result4} Snapshots of the vortex filaments in the plane perpendicular to  the counterflow for BS simulations, from Fig.~\ref{result1} (right), each at $t=10\,$s. 
The left panel is for the simulation with the Type II reconnection algorithm, Type III (middle) and Type IV ($\epsilon=1)^{-4}$ (right).}
  \end{center}
\end{figure*}

The results for the initial simulations using both the LIA and the full BS law are presented in Figs.~\ref{result1}--\ref{result4}.
Figure \ref{result1} shows the vortex line density plotted as a function of time for the LIA (left) and the BS (right) simulation, using various reconnection algorithms.
The results for the Biot-Savart law are very encouraging showing little difference between the simulations with the different reconnection algorithms.
In all simulations there is an initial rapid growth in the vortex line density, followed by a more gradual increase until eventual saturation to a fluctuating steady state.

As we discussed earlier, the reconnection method is the one ad-hoc aspect of the VFM, and it seems that, at least for quantum turbulence driven by counterflow, the results are very robust.
Note infact that all calculations performed with the BS law (Fig.~\ref{result1} right) converge to approximately the same value of $L$.
We believe this is a very useful result as this means that one can confidently draw conclusions from simulations which use the BS method.
We note, however, that one should probably always err on the side of caution with VFM simulations and check the effect of different reconnection algorithms with a different numerical experiment.

Figure \ref{result2} (right) shows the number of reconnections per unit time per unit length, $\zeta/\Lambda$, for the BS simulations, and the picture is the same.
For each of the different algorithms the reconnection rate per unit line length is approximately the same. The only exception is the simulation with the Type I reconnection algorithm, which shows a slightly higher reconnection rate.
This is understandable as this is the algorithm which has the least restrictions and simply reconnects filaments if the distance between them is less than $\delta/2$; however there is no discernible impact on the evolution of the vortex line density.
Snapshots of the system at the end of simulations are plotted in Fig.~\ref{result4}.
We shall return to discuss the further simulations using the BS method at the end of this section.

The picture for the LIA simulations is very different. We find that even with the same initial conditions, slight differences in the reconnection algorithms lead to large changes in the overall evolution of the system.
Kondaurova and Nemirovskii \cite{Konda2008} found that using a reconnection scheme similar to the Type IV algorithm used here they reached a steady state.
We find that certain simulations do reach a steady state, although the steady states we find are degenerate.
By degenerate we mean that the system evolves to a set of very straight vortices, arranged in planes parallel to the counterflow direction.
These planes are then simply advected in the direction of the counterflow with a negligible change in vortex line density.
In this configuration the reconnection rate drops to zero.
This is the eventual fate of the simulations with Type I, II and IV ($\epsilon=10^{-3},10^{-2}$) algorithms.

In the other simulations a steady state is not found and large fluctuations in the line density are seen.
We find the same bundles of vortex lines, stratified in layers, observed by Schwarz \cite{Schwarz} and recently reproduced by Adachi \cite{Adachi2010}.
Schwarz recognised that this is a spurious effect and introduced his artificial mixing procedure to avoid it \cite{Schwarz}, see Fig.~\ref{result3} (right).
The reconnection rate per unit line length, Fig.~\ref{result2} (left), also show that LIA simulations are highly sensitive to the reconnection algorithm used.
We thus agree with other authors that, due to the absence of the non-local component of the velocity field, LIA is unsuitable for simulations of turbulence\cite{Adachi2010}.

\begin{figure}
  \begin{center}
    \psfrag{L}{$\sqrt{L}$}
    \psfrag{v}{$v_{ns}$}
    \includegraphics[width=0.47\textwidth]{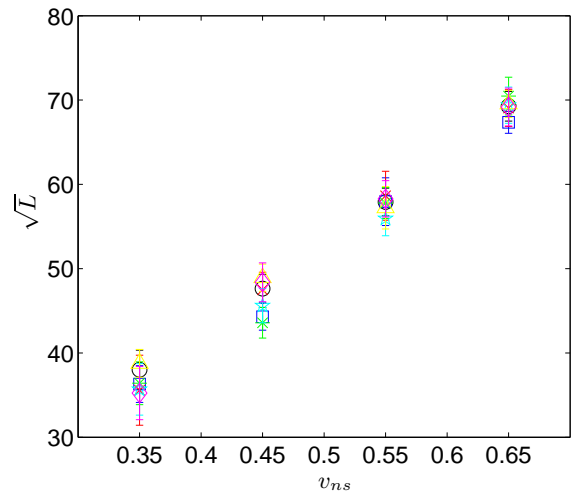}
    \caption{\label{result5} A plot of the root of the vortex line density, $L$, and the magnitude of the counterflow velocity, $v_{ns}$. The linear relationship between $\sqrt{L}$ and $v_{ns}$ for each of the reconnection schemes is apparent, plotting symbols and colors is as in Fig.~\ref{result1}.}
  \end{center}
\end{figure}

We now return to the BS simulations, focusing on the affect of varying the counterflow velocity $v_{ns}$.
We are interested in the relationship between the steady state line density and the counterflow velocity.
As above, the temperature is fixed at 1.6K
Figure \ref{result1} (right) is very representative of the time series plot for the line density, at different temperatures, and so we do not reproduce further plots for each of the different values of $v_{ns}$.
Instead in Fig.~\ref{result5} we display the linear relationship between $\sqrt{L}$ and $v_{ns}$, according to Eq.~\ref{eq:Ldensity}.
Again, we find that the result is not sensitive to which reconnection algorithm is used.

Finally it is instructive to compare the effect of the reconnection algorithm used on the scaling parameter, $\gamma$,
and compare with experimental results.
Our aim here is not to claim that the algorithm which yields the value closest to that experimental value should be used in the VFM.
Instead we wish to add further weight to our claims that results from the VFM can be informative, so long as a physically justifiable reconnection algorithm is used.
Table 1. shows the values of $\gamma$ for the different reconnection algorithms. 
The experimental value of $\gamma$ at this temperature is 93 \cite{Tough,Childers} ; Adachi et al. \cite{Adachi2010} gave numerical results of $\gamma=109.6$.

\begin{table}
\begin{center}
\begin{tabular}{| c | c |}
  \hline
  reconnection algorithm & $\gamma$\\
  \hline                       
  Type I & 100.6  \\
  Type II & 111.5  \\
  Type III & 111.7 \\
  Type IV ($\epsilon=10^{-4}$) & 104.1  \\
  Type IV ($\epsilon=10^{-3}$) & 106.8  \\
  Type IV ($\epsilon=10^{-2}$) & 111.6  \\
  Type IV ($\epsilon=10^{-1}$) & 116.4  \\
  \hline 
\end{tabular}
\caption{A table of values of $\gamma$ for the different reconnection algorithms in simulations using the Biot-Savart law at T=1.6K} 
\end{center}
\end{table}

\section{Conclusions}\label{sec:conclusions}

In conclusion we have shown that the vortex filament method is very robust to the reconnection algorithm used, provided one uses the full Biot-Savart integral to determine the velocity field.
In contrast results from simulations using the local induction approximation are highly dependent on the algorithm used, lending further weight to the criticism of this method.

\section{Acknowledgements}

We thank C. F. Barneghi for useful discussions.
This work was supported by the  Leverhulme Trust.

\end{document}